# DAΦNE OPERATING EXPERIENCE WITH CRAB WAIST COLLISIONS

M. Zobov, INFN Laboratori Nazionali di Frascati, Frascati, Italy
for DAΦNE Collaboration Team[1]

*Abstract*

The Φ-factory DAΦNE was upgraded in the second half of 2007 in order to implement a recently proposed scheme of crab waist collisions aimed at substantial luminosity increase. Commissioning of the modified collider started in November 2007. In this paper we briefly describe the crab waist collision concept and discuss in detail the DAΦNE hardware upgrade and obtained experimental results.

## INTRODUCTION

DAΦNE is an electron-positron collider working at the c.m. energy of the Φ resonance (1.02 GeV c.m.) to produce a high rate of K mesons [2]. In its original configuration the collider consisted of two independent rings having two common Interaction Regions (IR) and an injection system composed of a full energy linear accelerator, a damping/accumulator ring and transfer lines. Figure 1 shows a view of the DAΦNE accelerator complex while some of the main collider parameters are listed in Table 1.

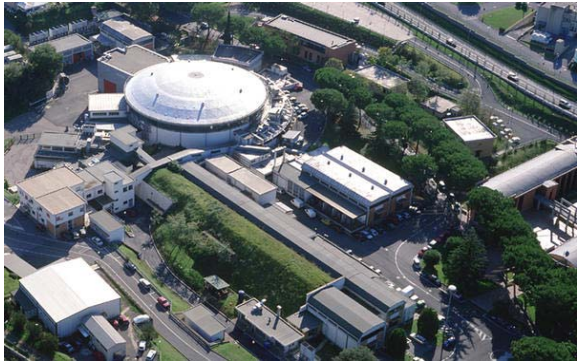

Fig. 1 View of DAΦNE accelerator complex.

Since year 2000 DAΦNE has been delivering luminosity to three experiments, KLOE [3], FINUDA [4] and DEAR [5], steadily improving performances in terms of luminosity lifetime and backgrounds. In these years the collider has undergone several progressive upgrades implemented during the shutdowns for detector changeover. The best machine performances were obtained in the KLOE and FINUDA runs. In particular, we reached a peak luminosity of $1.6 \times 10^{32}$ cm$^{-2}$s$^{-1}$ with a maximum daily integrated luminosity of about 10 pb$^{-1}$.

Recently the Crab Waist (CW) collision scheme has been proposed for collider luminosity increase [6, 7]. It holds the promise of substantial increasing the luminosity of storage-ring colliders beyond the state-of-the-art, without any significant increase in beam current and without reducing the bunch length. This scheme was adopted in 2007 for the design of the new interaction region of DAΦNE for the SIDDHARTA experiment [8].

## CRAB WAIST CONCEPT

The Crab Waist (CW) scheme of beam-beam collisions can substantially increase the luminosity of a collider since it combines several potentially advantageous ideas. Let us consider two bunches colliding under a horizontal crossing angle $\theta$ (as shown in Fig. 1a). Then, the CW principle can be explained, somewhat artificially, in the three basic steps (see [7] for details):

**1 Step**: Large Piwinski Angle (LPA) ($\Phi = 0.5\theta\sigma_z/\sigma_x$) at the interaction point (IP), see Fig. 1a. It is obtained by increasing the beam crossing angle and by reducing the transverse horizontal beam size $\sigma_x$. The LPA allows reducing the beam overlap length proportionally to $\sigma_x/\theta$. The luminosity gain can be achieved by increasing the bunch current. With the large $\Phi$ this becomes possible due to the fact that the vertical tune shift $\xi_y$ decreases proportionally to $(1/\Phi)^{1/2}$, while the horizontal tune shift $\xi_x$ drops even faster, $\sim(1/\Phi)$ [9]. Moreover, the LPA reduces the detrimental effects of parasitic crossings (PC) because of the higher crossing angle and smaller horizontal beam size.

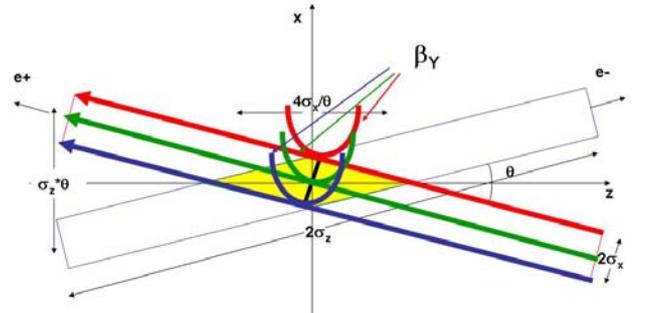

a) Crab sextupoles OFF.

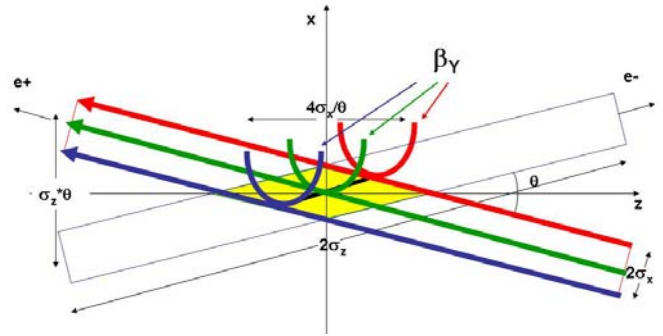

b) Crab sextupoles ON.

Figure 1: Crab Waist collision scheme.

**2 Step**: vertical beta function $\beta_y$ at the IP of the same order of the overlap area length $2\sigma_x/\theta$. Together with the reduction of the horizontal beam size, this is the main source of the geometric luminosity increase. Besides, it allows reducing the vertical tune shift [10].

**3 Step**: CW transformation rotating the vertical beta function of colliding beams as shown in Fig.1b. It is provided by sextupole magnets placed on both sides of the IP in phase with the IP in the horizontal plane and at $\pi/2$ in the vertical one (as shown in Fig. 2).

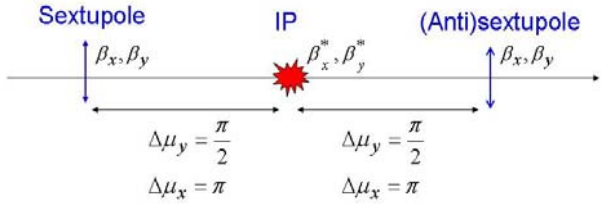

Figure 2: Crab sextupole locations.

The crab sextupole strength should satisfy the following condition depending on the crossing angle and the beta functions at the IP and the sextupole locations:

$$K = \frac{1}{\theta}\frac{1}{\beta_y^*\beta_y}\sqrt{\frac{\beta_x^*}{\beta_x}} \qquad (4)$$

The crab waist transformation gives a small geometric luminosity gain due to the vertical beta function redistribution along the overlap area. It is estimated to be of the order of several percent. However, the dominating effect comes from the suppression of betatron (and synchrobetatron) resonances arising (in collisions without CW) due to the vertical motion modulation by the horizontal betatron oscillations [11].

## HARDWARE UPGRADE

In 2007, during a five month shutdown used for the installation of the experimental detector SIDDHARTA, DAΦNE was upgraded implementing the new collision scheme based on large Piwinski angle and crab waist. This required important changes in the design of the mechanical and magnetic layout of the both interactions regions. Besides, several hardware modifications have been incorporated in order to improve the collider performance. Table 1 shows a comparison of the design beam parameters for the DAΦNE upgrade with those of the previous FINUDA run.

### Experimental Interaction Region

As it is seen in Fig.1, removing the splitter magnets and rotating the two sector dipole magnet in the long and short arcs adjacent to the interaction regions of both rings has doubled the horizontal crossing angle. Four additional corrector dipoles are used to match the vacuum chambers in the arcs.

The low-beta setion in the SIDDHARTA experimental interaction region is based on permanent magnet quadrupole doublets (see Fig. 3). The first quadrupole is defocusing and is shared by the two beams. Due to the off-axis beam trajectory it also provides strong beam separation. The second quadrupole, the focusing one, is installed just after the ring's beam pipe separation and is therefore on axis.

Table 1. Comparison of beam parameters for FINUDA run and for DAΦNE Upgrade.

| | DAΦNE FINUDA | DAΦNE Upgrade | |
|---|---|---|---|
| $\theta_{cross}/2$ (mrad) | 12.5 | 25 | |
| $\varepsilon_x$ (mm×mrad) | 0.34 | 0.20 | Larger Piwinski angle |
| $\beta_x^*$ (cm) | 170 | 20 | |
| $\sigma_x^*$ (mm) | 0.76 | 0.20 | |
| $\Phi_{Piwinski}$ | 0.36 | 2.5 | |
| $\beta_y^*$ (cm) | 1.70 | 0.65 | Lower vertical beta |
| $\sigma_y^*$ (μm) | 5.4 (low current) | 2.6 | |
| Coupling, % | 0.5 | 0.5 | |
| $I_{bunch}$ (mA) | 13 | 13 | |
| $N_{bunch}$ | 110 | 110 | |
| $\sigma_z$ (mm) | 22 | 20 | |
| L (cm$^{-2}$s$^{-1}$) ×10$^{32}$ | 1.6 | 5 | |

It worth mentioning that the new configuration almost cancels the problem of parasitic beam-beam long-range interactions. The two beams experience only one parasitic crossing inside the defocusing quadrupole where, due to the large horizontal angle, they are already well separated (about 20 $\sigma_x$).

The crab-waist sextupoles are installed at both ends of the interaction region. They are electomagnets with an integrated gradient about a factor of 5 higher than that of the normal sextupoles used for chromaticity correction.
Four electromagnetic quadrupoles have been installed on both sides of the experimental interaction region to provide the proper phase advance between the crab-waist sextupoles and the interaction point.

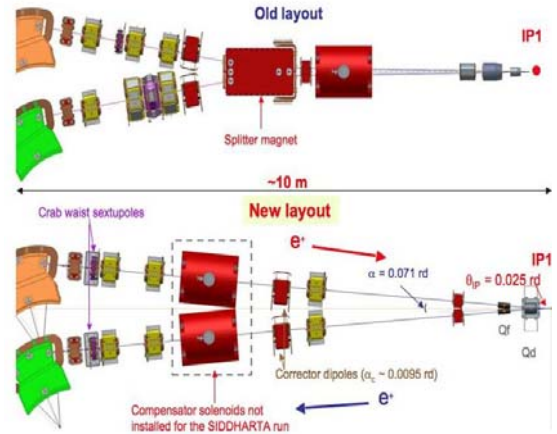

Fig.3 Half view of the old (top) and new (bottom) interaction region layout.

The compensator solenoids, that were present in the original setup, have been removed since there is no solenoid around the SIDDHARTA detector.

While designing the interaction region much attention has been paid to avoid any sharp discontinuities to keep the ring beam coupling impedance low. The vacuum chamber is rather simple, it consists of straight pipes merging in a Y-shape section. Special care has been given to the Y-section design in order to minimize power losses due to trapped high order modes (HOM). Numerical simulations have shown that even in the worst case the released power should not exceed 200 W. Nevertheless, the section has been equipped with cooling pipes to remove heating due to the HOMs, if necessary.

*Crossing region modification*

A new section providing complete separation of the two beams has replaced the second interaction region. It is geometrically symmetric to the experimental interaction region. Its independent beam vacuum chambers have been designed by splitting the original round pipe in two half-moon shaped sections, as seen in Fig. 4, thus providing full vertical beam separation. This aspect is quite important because such a separation completely cancels parasitic electomagnetic long-range beam-beam interaction.

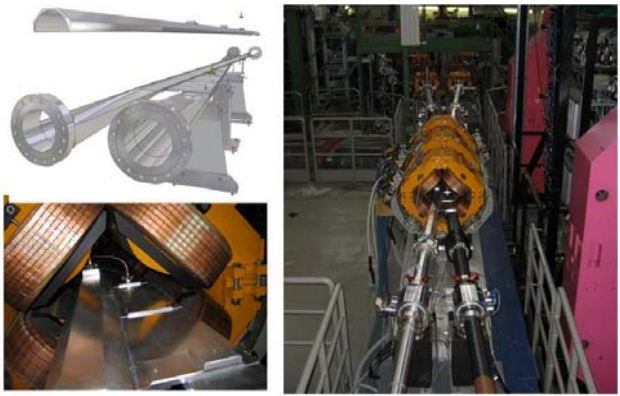

Fig.4 Crossing region view.

The magnetic layout of the crossing section is similar to that of the experimental interaction region, except for missing crab waist sextupoles and the central focusing region. Since there is no need to obtain low betas in the section the permanent quadrupoles doublet is substituted by a large aperture elecromagnetic quadrupole triplet allowing for wide operation flexibility in terms of betatron functions.

*New bellows*

New bellows have been developed and installed in the new interaction regions [12]. Their innovative component is the RF shield necessary to avoid the pipe discontinuity acting as a cavity for the beam. The new RF shield is implemented by means of $\Omega$ shaped Be-Cu strips, installed all around two cylindrical shells fixed at the bellows ends.

HFSS simulations in the frequency domain from DC to 5 GHz have shown that the new design reduces bellows contribution to the ring beam coupling impedance.

*New fast injection kickers*

The injection kickers, two in each main ring, have been replaced with new devices [13] based on tapered strips embedded in a rectangular cross section vacuum chamber allowing injection rate up to 50 Hz. The deflecting field is provided by the magnetic and the electric fields of a TEM wave travelling in the kicker structure, which generates short pulses, perturbing only 3 bunches out of 110 usually colliding. This new injection scheme represents a relevant improvement with respect to the old one, which was based on injection kickers having 150 ns pulse length perturbing almost half of the bunch train. Moreover, a smooth beam pipe and tapered transitions reduce the kicker contribution to the total ring coupling impedance budget. All these features are aimed at increasing the maximum storable currents, improving colliding beam stability and detector background reduction.

*Other upgrades*

Several other important upgrades and innovations should be also mentioned. Among them:
- commissioning the new luminosity monitor consisting of three different devices [14];
-upgrade of the DAΦNE control system;
-feedback system upgrade [15];
-removal of ion clearing electrodes;
- repositioning of several magnetic elements aimed at machine optics flexibility improvement and dynamic aperture optimization.

## NEW SCHEME COMMISSIONING

DAΦNE operation restarted at the end of November 2007 with the goal to optimize the collider performance, test new hardware, verify the crab waist collision concept and provide luminosity to the SIDDHARTA experiment.

*Main ring optics*

In the early stage of the commissioning we used a detuned optics, with larger beta functions at the interaction point, in order to simplify beam injection, put the diagnostics in operation and perform a satisfactory optics modeling.

Beam closed orbit has been minimized together with the steering magnet strengths relying also on beam based procedure to point out and fix misalignment errors. Vertical dispersion has been minimized by global vertical orbit correction and by centering the beam vertical position in the arc sextupoles.

Once a reliable machine model has been defined, the ring optics has been moved progressively towards the nominal one. Now, as shown in Fig.5, the measured optical functions are in a good agreement with the model predictions. The present beta functions at the interaction point $\beta^*_{y/x} \sim 0.0085/0.25$ m are only slightly larger than the design values ($\beta^*_{y/x} \sim 0.0065/0.20$ m).

The transverse betatron coupling has been minimized mainly by correcting rotation errors in the low-β horizontally focusing quadrupoles, now independent for the two rings. The best value obtained is better than $\kappa = 0.2\%$ for both beams measured at the synchrotron light monitor. The low coupling together with the low beta function helped to obtain the smallest vertical beam size at the IP (3.5 μm) ever measured in DAΦNE.

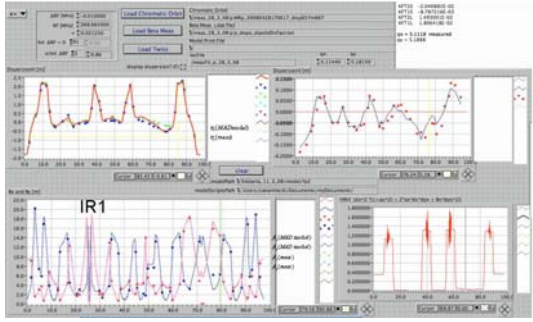

Fig.5 DAΦNE optical functions: dispersion, second order dispersion, beta functions (Solid lines – optics model, dots – measurements)

In order to be able to switch on the crab waist sextupoles, that are much stronger than the ordinary ones used for chromaticity correction, beam orbit deviations have been carefully corrected to avoid tune changes. Two electromagnetic quadrupoles have been added symmetrically with respect to the IP in each ring in order to finely tune the phase advances between the crab waist sextupoles and the IP.

The nonlinear optics has been optimized by tuning the sextupole configuration to minimize the dependence of the betatron functions on energy and to improve the dynamic aperture and, therefore, the beam lifetime.

*High current issues*

The ring beam coupling impedance is responsible for several harmful effects in beam dynamics affecting also the collider luminosity, as observed during previous DAΦNE runs [16]. The careful design of the upgraded vacuum chamber, including smoother interaction region vacuum pipes, new bellows, new injection kickers, removal of ion clearing electrodes etc., has allowed reducing the coupling impedance of the positron ring by about 50% and that of the electron one by approximately 70% [17]. One can see in Fig.6 that now the bunch length in both rings does not exceed 2 cm till 15 mA/bunch, as required by the design specifications.

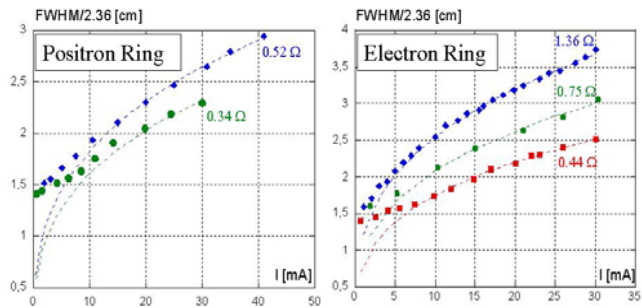

Fig.6 Bunch lengthening in the positron (left) and the electron rings (right)

Multibunch high current operation strongly depends on vacuum conditions. The new vacuum chambers have required a time consuming beam conditioning. After three months of operation a reasonable vacuum condition has been obtained. However, no more than 95 electron bunches can be stored, so far, in collision instead of the 110 used during the past DAΦNE runs.

In turn the positron beam is affected by a fast transverse instability leading to beam losses. Such instability has been cured partially by tuning the transverse and longitudinal feedback systems. The four e- and e+ transverse feedbacks have been upgraded by adopting the new iGP (Integrated Gigasample Processor) feedback unit [15]. Beyond its ordinary stabilization function, this system allows building a variety of diagnostic tools ranging from the single bunch tune to the single turn beam position measurement, a feature which has been extensively used in the injection induced transient analysis aimed at reducing the impact of the injection kickers on the maximum storable current, especially for the e+ beam.

A further e+ current improvement has been obtained by installing solenoid windings ($B_{sol} \sim 45$ Gauss) in the long sections of the IR and of the RCR in the e+ ring. Solenoids have been effective especially in the first commissioning phase; in fact they reduce the transverse instability rise-time boosting the action of the transverse feedbacks.

A relevant contribution is expected when the fast high voltage pulsers (5 ns, 40kV), perturbing only three bunches during injection, will replace the old ones (200 ns, 25kV) presently feeding the new stripline kickers.

As an overall result the highest currents stored by now are in single beam: I- = 1.8 A (95 bunches), I+ = 1.15 A (120 bunches) and in 95 colliding bunches: I- = 1.2 A, I+ = 1.1 A.

## EXPERIMENTAL RESULTS

Several measurements and qualitative observations of the beam-beam behavior have confirmed the effectiveness of the new collision scheme.

The simplest and most evident test consists in switching off the crab waist sextupoles of one of the colliding beams. As a consequence, the blow up of both horizontal and vertical transverse beam sizes of that beam can be observed (see Fig.7) together with a luminosity reduction recorded by all the luminosity monitors. Such a behavior is compatible with the appearance of beam-beam resonances when the crab sextupoles are off, as predicted.

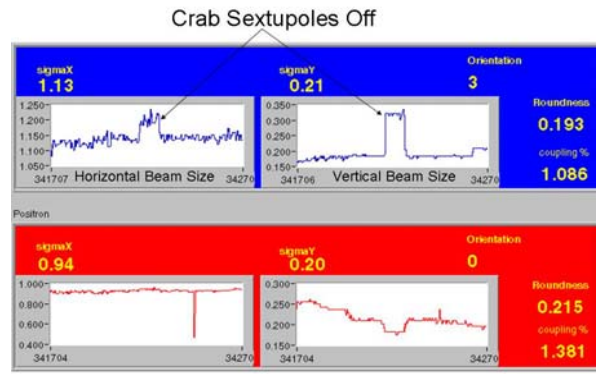

Fig.7 Transverse beam sizes at the synchrotron light monitors (electrons - blue, positrons - red).

The specific luminosity, defined as the luminosity divided by the product of the total currents in the two beams (taking into account also the number of colliding bunches), exceeds by 3÷4 times the best value measured during the past DAΦNE runs, and the beam-beam tune shift exhibits a fairly linear behaviour as a function of current per bunch in the opposite colliding beam.

In order to avoid multibunch effects that can affect the luminosity, such as ion trapping, e-cloud, gap transients etc., we have studied collisions with a small number of bunches. In particular, Fig. 8 shows the case of 10 colliding bunches per beam. As it is seen, the maximum achieved luminosity per bunch is about $4 \times 10^{30} cm^{-2}s^{-1}$. This value is consistent with the single bunch luminosity predicted by numerical simulations taking into account the present optics parameters at the IP and the strength of the crab waist sextupoles. Extrapolating this result to 110 colliding bunches would yield a luminosity close to the design one.

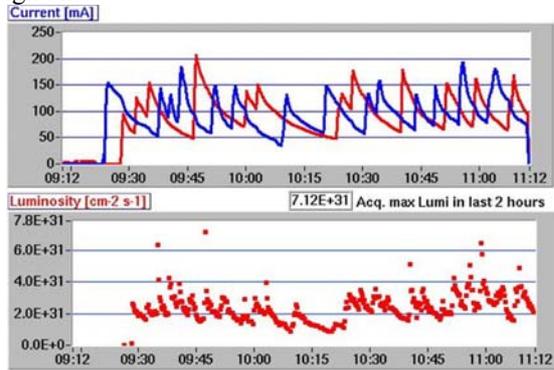

Fig.8 Beam currents (upper plot) and corresponding luminosity (lower plot) for 10 colliding bunches.

We have performed also dedicated multibunch tests colliding 95 bunches per beam for several hours with the crab waist sextupoles on and off. The clear improvement in luminosity with the CW sextupoles on is seen in Fig. 9.

The most relevant results of the commissioning concern the luminosity. So far the maximum measured peak luminosity is in excess of $L_{peak} = 2.2 \ 10^{32} \ cm^{-2}s^{-1}$ (see Fig. 10), the best daily integrated luminosity is $L_{\int day} \sim 8 \ pb^{-1}$ and the highest integrated luminosity in one hour is $L_{\int 1hour} \sim 0.5 \ pb^{-1}$ averaged over a two hours run

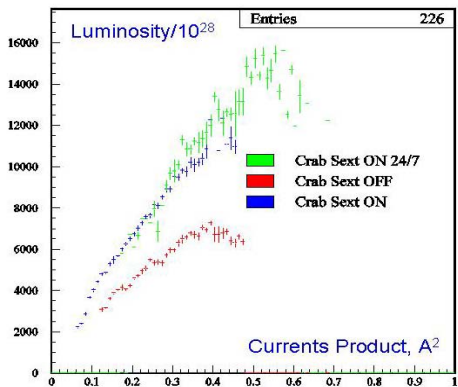

Fig.9 DAΦNE peak luminosity with crab sextupoles off (lower curve) and on (upper curves).

Further luminosity increase is expected from fine machine tuning with lower vertical beta function at the IP and fully exploiting recently installed stronger crab waist sextupoles. Moreover, the number of colliding bunches, due to the recovered vacuum conditions and to the recent developments on the transverse and longitudinal feedback systems, can be raised from 95 to 110, as during the past DAΦNE runs. The average luminosity can also profit from speeding up the switching procedure between electron and positron injection. With all these improvements the peak luminosity is expected to reach $4 \times 10^{32} \ cm^{-2}s^{-1}$ and monthly integrated luminosity $\sim 0.5 fb^{-1}$.

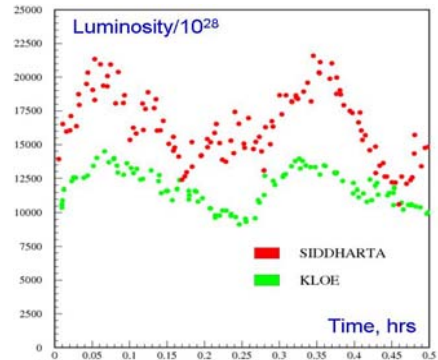

fig.10. DAΦNE luminosity over 2 hours for KLOE run (green) and for crab waist scheme (red).